\documentstyle[prl,twocolumn,psfig,aps]{revtex}

\newcommand{\be}{\begin{equation}}
\newcommand{\ee}{\end{equation}}
\newcommand{\bea}{\begin{eqnarray}}
\newcommand{\eea}{\end{eqnarray}}
\newcommand{\no}{\noindent}

\newcommand{\sss}{\scriptscriptstyle}
\newcommand{\stb}{\;\raisebox{-.5ex}{\rlap{$\sss PT$}}
\raisebox{.3ex}{$\longrightarrow$}\;}

\newlength{\figwidth}
\newlength{\figheight}
\begin{document}
\draft

\title{Mesons Above The Deconfining Transition}

\author{{\it QCD-TARO Collaboration:}
 Ph.~de~Forcrand$^1$,
 M.~Garc{\'\i}a~P\'erez$^2$,
 T.~Hashimoto$^3$,
 S.~Hioki$^4$,
 H.~Matsufuru$^{5,6}$,
 O.~Miyamura$^6$,
 A.~Nakamura$^7$,
 I.-O.~Stamatescu$^{5,8},$
 T. Takaishi$^9$
 and T.~Umeda$^6$}

%\author{(QCD-TARO Collaboration)\bigskip}

\address{$^1$ETH-Z\"urich, CH-8092 Z\"urich, Switzerland}
\address{$^2$Dept. F\'{\i}sica Te\'orica, Universidad Aut\'onoma de Madrid,
          E-28049 Madrid, Spain}
\address{$^3$Department of Applied Physics, Faculty of Engineering,
                              Fukui University, Fukui 910-8507, Japan}
\address{$^4$Department of Physics, Tezukayama University,
                Nara 631-8501, Japan}
\address{$^5$Institut f\"ur Theoretische Physik, Univ. Heidelberg
           D-69120 Heidelberg, Germany}
\address{$^6$Department of Physics, Hiroshima University,
                                  Higashi-Hiroshima 739-8526, Japan}
\address{$^7$Res. Inst. for Information Science and Education, Hiroshima
          University, Higashi-Hiroshima 739-8521, Japan}
\address{$^8$FEST, Schmeilweg 5, D-69118 Heidelberg, Germany}
\address{$^9$Hiroshima University of Economics, Hiroshima 731-01, Japan}

\date{\today}
\maketitle

\vspace{-0.5cm}

\begin{abstract} We analyze {\it temporal}  and
{\it spatial} meson correlators in quenched lattice $QCD$
at $T>0$. Above $T_c$
we find different masses and  (spatial) ``screening masses", signals of
plasma formation, and indication of persisting
``mesonic" excitations.
\end{abstract}

\pacs{PACS numbers: 12.38.Gc, 12.38.Mh}

%\narrowtext
\vspace{-0.5cm}

\no With increasing temperature we expect the physical picture of
QCD to change according to a phase transition where
chiral symmetry restoration
and deconfinement may  simultaneously occur.
For model independent non-perturbative  results one attempts
lattice Monte Carlo studies \cite{edw}.
Since in the Euclidean formulation the $O(4)$ symmetry is
broken at $T>0$ (see, e.g., \cite{tgen}),
 physics appears different, depending on whether we probe the space 
(``$\sigma$": ${\bf x}$) or time  (``$\tau$": $t$)
 direction: the string tension, e.g.,
 measured from $\sigma\sigma$  Wilson loops
does not vanish above $T_c$, in contrast to the one measured from
 $\sigma\tau$  loops.
{\it Therefore we need to investigate hadronic
correlators with full ``space-time" structure, in particular the
propagation in the Euclidean time}.  The latter, however, poses
special problems because of the inherently limited
{\it physical length} of
the lattice in the time direction
$ l_{\tau}= 1/T$. We shall  discuss 
 briefly this question and introduce our procedure.

1) Lattice problems: Large $T$ can be
achieved  using small
$N_{\tau} = l_{\tau}/a$ ($a$: lattice spacing), 
however this leads to systematic errors \cite{fthl}.
Moreover, having the
$t$-propagators at only a few points makes it difficult
to characterize the unknown structure in the corresponding channels.
To  obtain a fine
$t$-discretization
and thus  detailed
 $t$-correlators, while avoiding
prohibitively large lattices (we need large spatial size to
avoid finite size effects, typically $l_{\sigma} \sim 3 l_{\tau}$),
we use
different lattice spacings
in space  and in time,
$a_{\sigma}/a_{\tau}=\xi >1 $.
For this we employ anisotropic
 Yang-Mills and fermionic actions \cite{FKa}:
\bea
&&S_{\rm YM}=-(\beta/3)(\gamma_G^{-1}\hbox{ReTr}
\Box_{\sigma\sigma}+\gamma_G \hbox{ReTr}
\Box_{\sigma \tau})
\label{e.act}\\
&&S_{\rm F}=(2 \kappa_{\sigma})^{-1} {\bar{\Psi}} W \Psi,\ \
\kappa_{\sigma}^{-1} =
     2(m_0 +3 + \gamma_F),\\
&&W=1 -
\kappa_{\sigma}\ (\
\mbox{
$\sum_i$ }
\Gamma_i^+ U_i T_i +
\gamma_F\Gamma_4^+ U_4 T_4
+\hbox{``h.c."}\
)\label{e.fact}
\eea
\no  ($\Gamma_{\mu}^{\pm} =
1 \pm \gamma_{\mu}$, $\gamma_{\mu}^2 = 1$, $T_{\mu}$ are
 lattice translation operators \linebreak and $m_0$ the bare quark mass).
 $ \xi$  is determined from $T \simeq 0$ correlators
$F$ (``calibration") by requiring isotropy in the physical distance: 
$F^{\sigma}(z) = F^{\tau}(t = \xi z)$.
In a quenched simulation at some $\gamma_G$, $\xi$ is fixed 
 by the
Yang-Mills calibration, then
$\gamma_F$ is tuned to give the same $\xi$
for $T \simeq 0$ hadron correlators.
We vary $T = \xi / N_{\tau} a_{\sigma}$ by varying $N_{\tau}$.

2) Physical problems:
Increasing the temperature is expected to induce
significant changes in the structure of the hadrons (see, e.g. \cite{hild} for
reviews). Two ``extreme" pictures are
 frequently used for the
intermediate and the high $T$ regimes, respectively:
the weakly interacting
meson gas, where we expect the
mesons to become effective resonance modes with a small mass shift
and width due to
the interaction; and the quark-gluon plasma (QGP), where
 the mesons should eventually disappear
and at very high $T$ perturbative effects should dominate.
These genuine temperature effects should be reflected in the
low energy structure in the mesonic channels. But this structure
cannot be observed directly,
due to the inherently coarse {\it energy} resolution $1/l_{\tau}=T$.
Our strategy is the following: we  fix
at $T=0$ a mesonic source which gives a large (almost 100\%)
projection onto the
ground state. Then we use this source to determine the changes induced by the
temperature on the ground state. For $T>T_c$
we do not have a good justification to use
 that source as representative of the meson but we assume that
it still projects onto the dominant
low energy structure in the
spectral function. This is a reasonable procedure if
 the mesons interact weakly with other
hadron-like modes in the thermal bath and
 the changes in the correlators
are small. Large changes  will signal the
breakdown of this weakly interacting gas picture  and there
we shall  try to compare our observations
with the QGP picture.

On a {\it periodic} lattice the contribution of a pole 
in the mesonic spectral function to the $t$-propagator
is cosh$(m(t-N_{\tau}/2))$ (this $m$ is therefore called ``pole-mass"). 
A broad structure or the admixture of
excited states leads to a superposition of such terms. Fitting
a given $t$-propagator by  cosh$(m(t)(t-N_{\tau}/2))$ at pairs of points $t,t+1$
defines an ``effective mass" $m(t)$ which is constant
if the spectral function has only one, narrow peak.
We shall simply
speak about $m(t)$ as ``mass": it
connects directly to the (pole) mass of the mesons below $T_c$, while above
$T_c$ it will help  analyze the dominant low energy
structure in the frame of our strategy above.
By contrast,
we shall speak of ``screening mass" ($m^{(\sigma)}$)
when extracted from {\it spatial
propagators}. $m^{(\sigma)}$ is  different from the $T>0$  mass  
(the propagation in the space directions
represents  a $T=0$ problem with finite size effects).

\begin{figure}[tb]
\vspace*{-0.2cm}
\center{
\leavevmode\psfig{file=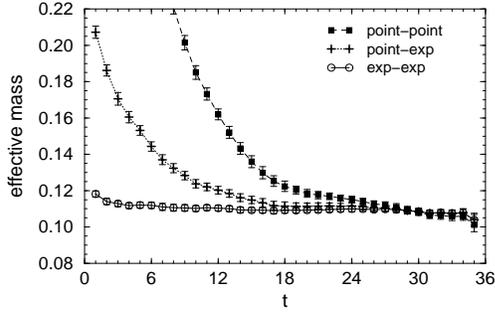,width=\figwidth}
}
\vspace{-0.5cm} \\
\caption{
Effective pseudoscalar mass $m(t)$ in units of $a_{\tau}^{-1}$
vs $t$, for various sources at $T \simeq 0$ ({\it Set-A}).}
\label{f.hem72}
\vspace{-0.2cm}
\end{figure}

We use lattices of $12^3 \times N_{\tau}$
with $N_{\tau}=72, 20,16$ and $12$ at $\beta = 5.68$, $\gamma_G=4$.
We find \cite{TARO96}
$T_c$  at $N_{\tau}$ slightly above 18, which fixes
for the  above lattices  $T \ \simeq 0, 0.93 T_c, 1.15 T_c$ and
$1.5T_c$ and $a_{\tau} \sim 0.044$ fm $=(4.5$ GeV$)^{-1}$.
We present two sets of results:
{\it Set-A} represents a prospective study of the $T-$dependence
of the {\it temporal} propagators,
 calibrated  with
$\xi = 5.9(5)$ (apparently overestimated); {\it Set-B} represents a 
more precise analysis of
the $T$ dependence of the {\it temporal} and {\it spatial}
correlators for 3 quark masses, with also a more precise
 calibration $\xi\simeq 5.3(1)$.
The various parameters are given in
the Table. Details will be given
in a forthcoming paper \cite{qcdt98}.
 We use periodic (anti-periodic) boundary
conditions in the spatial
(temporal) directions and gauge-fix to Coulomb gauge.
We investigate correlators of the form:
\vspace{-0.2cm}
\bea
G_M(x,t) =
\mbox{$\sum_{{\bf z},{\bf y_1},{\bf y_2}}$} w_1({\bf y_1})w_2({\bf y_2})
\times \hspace{2 cm} \nonumber \\
\langle {\rm Tr} \left[S({\bf y_1},0; {\bf z}, t) \gamma_M
\gamma_5 S^{\dag}({\bf y_2},0; {\bf z}+ {\bf x}, t) \gamma_5
\gamma_M^{\dag}\right]\rangle
\label{e.corr}\\
 w_{1,2}({\bf y}) \sim \delta({\bf y}) \ {\rm (}point{\rm )};\
 w_{1,2}({\bf y}) \sim {\rm exp}(-ay^p)\ {\rm (}exp.{\rm )}
\label{e.sour} 
\eea
\no with $S$ the quark propagator and $\gamma_M =
\{\gamma_5,  \gamma_1, 1, \gamma_1\gamma_5\}$ for
$M=\{Ps, V, S,
A\}$ (pseudoscalar, vector, scalar and
axial-vector, respectively). We use
point and smeared  (exponential) {\it quark} sources and point sink.
We fix the exponential ({\it exp}) source taking
the parameters $a,p$ in (\ref{e.sour}) from the observed
dependence on ${\bf x}$ of the temporal $Ps$ correlator  with
{\it point-point}
source at $T \simeq 0$ (see Table).  
The results of a variational analysis using
 {\it point-point},  {\it point-exp} and
{\it exp-exp} sources indicate that the latter ansatz projects practically
entirely onto the ground state at $T=0$. This is well seen from the
effective mass in Fig. \ref{f.hem72}.
Therefore we use throughout the {\it exp-exp} sources  in the 
Table, according to our
strategy for defining
hadron operators at high $T$ \cite{sour}. 
All masses are given in units  $a_{\tau}^{-1}$,
i.e. we plot $mass \times a_{\tau}$. Errors are statistical only.

\begin{figure}[tb]
\vspace*{-0.5cm}
\center{
\leavevmode\psfig{file=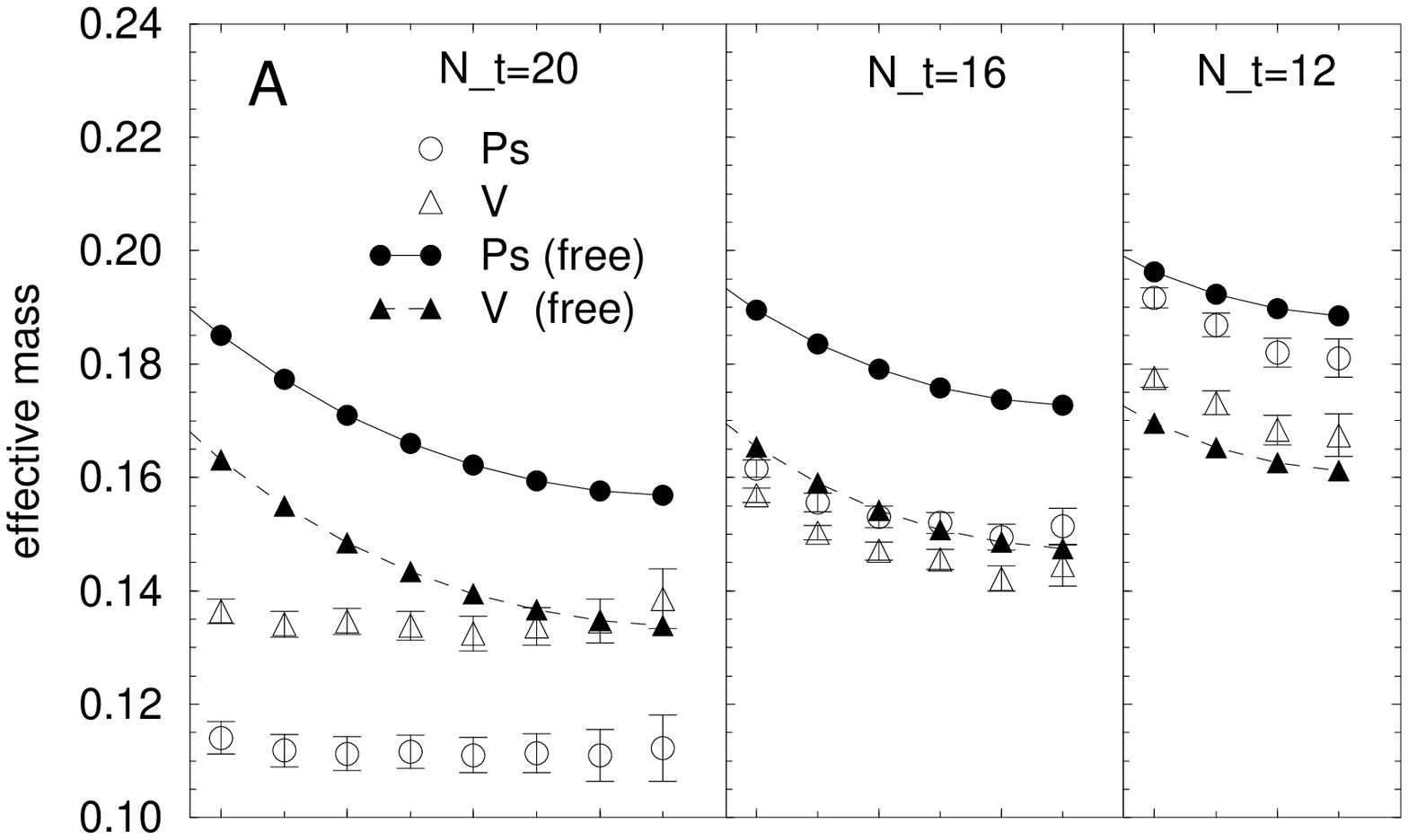,width=1.001\figwidth}
\vspace*{-1.10cm} \\
\hspace*{0.015cm}
\leavevmode\psfig{file=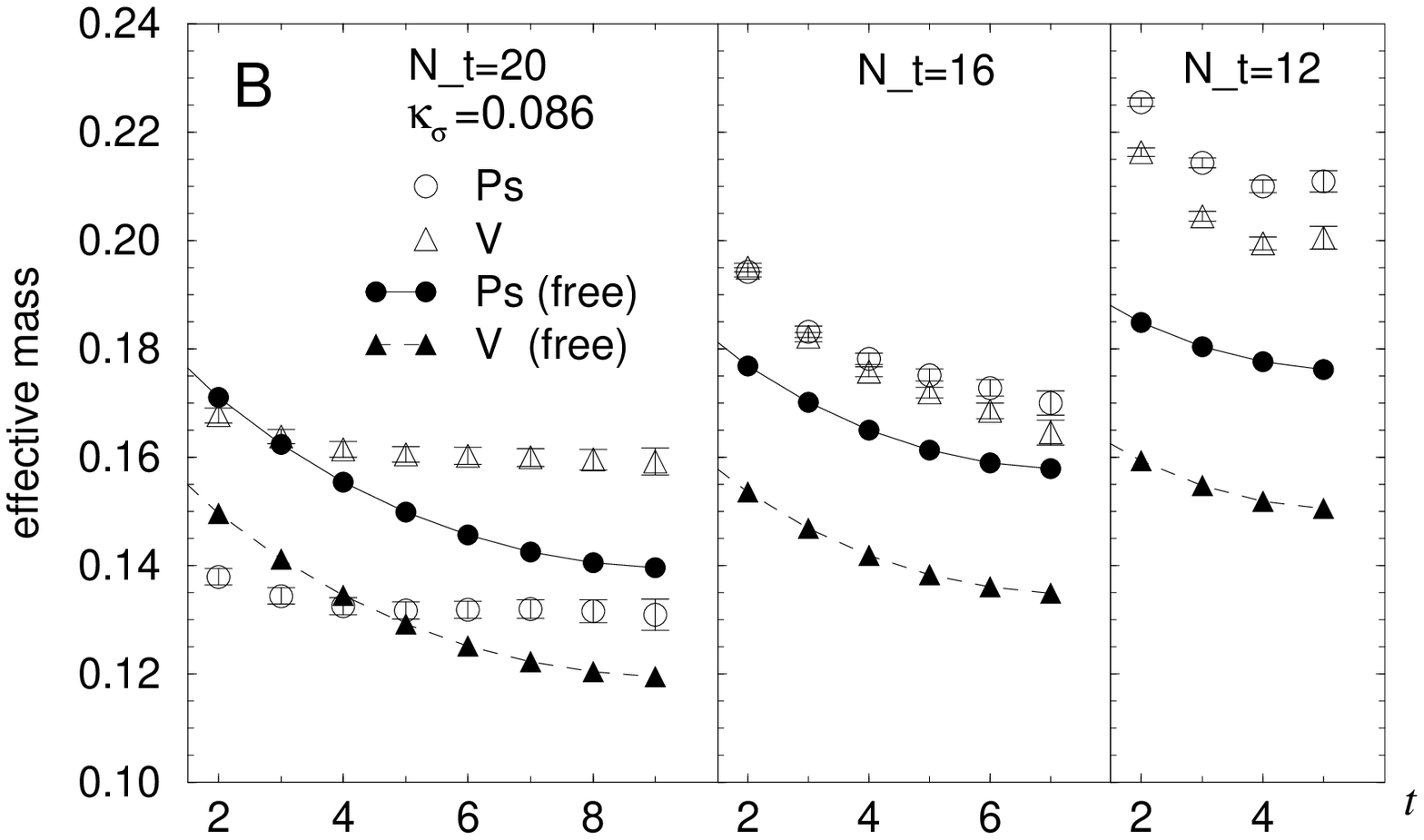,width=1.0\figwidth}
}
\vspace{-0.3cm} \\
\caption{
From left to right, effective  mass $m(t)$ in units of $a_{\tau}^{-1}$
at $T\simeq 0.93$, $1.15$ and $1.5T_c$ (open
symbols) vs $t$. Also shown are the effective masses
from the same correlators  calculated
using free quarks. A: {\it Set-A},
B: {\it Set-B}, $\kappa_{\sigma}=0.086$.
}
\label{f.hemT}
%\vspace{-0.5cm}
\end{figure}

{\it a) Effective masses.} In Fig. \ref{f.hemT}
 we show the effective  mass $m(t)$
of the $Ps$  and $V$ time-propagators at $T \simeq 0.93 T_c$,
$1.15T_c$ and $1.5 T_c$. The similarity
between the {\it Set-A} and {\it-B} data indicates that
calibration uncertainties are unimportant.
We notice practically no change from
$T=0$ (Fig. \ref{f.hem72})
to  $0.93 T_c$,
while above $T_c$ clear changes develop: 
$m(t)$ depends strongly on $t$, it increases
significantly, and
the $Ps$ and $V$ reverse their positions. 
Because of the large changes
 at high $T$  we compare here with
 the unbound quark picture of the high $T$ regime of QGP. 
For this we calculate mesonic
correlators using, with the same source,
free quark propagators  $S_0$ instead of $S$ in (\ref{e.corr}),
with  $\gamma_F=\xi=5.9(5.3)$ for
 {\it Set-A}({\it Set-B}) and, illustratively, 
$m_0 = m_q a_{\sigma}= m_q a_{\tau} \xi = 0.1$.
We did not attempt
a quantitative comparison at present
but similarities are apparent -- see Fig.  \ref{f.hemT}.
Generally above $T_c$,  $m(t)$ shows stronger $t$-dependence,
which means that the spectral function selected by
the {\it exp-exp} source no longer has just one, narrow contribution,
as for $T<T_c$.

\begin{figure}[tb]
\vspace*{-0.8cm}
\center{
\leavevmode\psfig{file=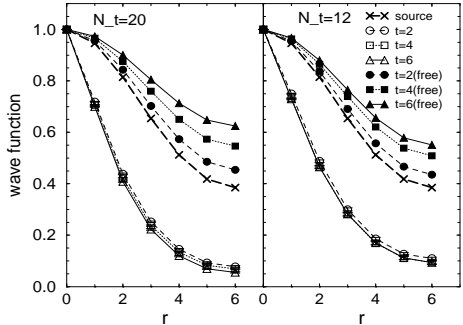,width=0.9\figwidth,height=0.8\figheight}
}
\vspace{-0.1cm} \\
\caption{
$Ps$ wave functions ({\it Set-B}, $\kappa_{\sigma}=0.086$,
 {\it exp-exp} source) 
normalized at $r=0$ 
vs quark separation $r$ at $t=2,4$ and $6$, using full and free propagators.
Also plotted is the initial distribution of separations as given by the source,
$ \int d^3y w({\bf y}+{\bf r})w({\bf y})$. $T \simeq 0.93T_c$ (left) and
$T \simeq 1.5T_c$ (right).
 }
\label{fig:wave_func}
%\vspace{-0.3cm}
\end{figure}

{\it b) Wave functions.} For the {\it Set-B} we have also analyzed
the ``wave functions", i.e. the behaviour of the temporal correlators with
 the quark-antiquark distance
at the sink.
In Fig. \ref{fig:wave_func} we compare the $Ps$
wave functions normalized at $x=0$,
$G_{Ps}(x,t)/G_{Ps}(0,t)$, at several $t$ for $T \simeq 0.93T_c$
(which is very similar to $T \simeq 0$) and $T \simeq 1.5T_c$  at our
lightest quark mass ($
\kappa_{\sigma}=0.086$)
 and for the free quark case
($m_q a_{\sigma}=0.1$, $\gamma_F=\xi = 5.3$).
Our {\it exp-exp} source 
appears somewhat too broad at $T \simeq 0.93T_c$:
the quarks go nearer each other while
propagating in $t$. Interestingly
enough, this is also the case at $T \simeq 1.5T_c$: the spatial distribution
 shrinks and stabilizes, indicating that even at this high
temperature there is a tendency for quark and anti-quark to stay together.
This is in clear contrast to the free quark case
 which never shows such a behaviour regardless of the source
(in Fig. \ref{fig:wave_func} we use the {\it exp-exp}
source and $m_0 = 0.1$; for heavy free quarks we
expect a ``wave function" similar to the source at all $t$). Hence the only
effect of the temperature on
the wave function is to make it slightly  broader.
The same holds also for
the other mesons at all quark masses.

\begin{figure}[tb]
\vspace*{0.25cm}
\center{
\leavevmode\psfig{file=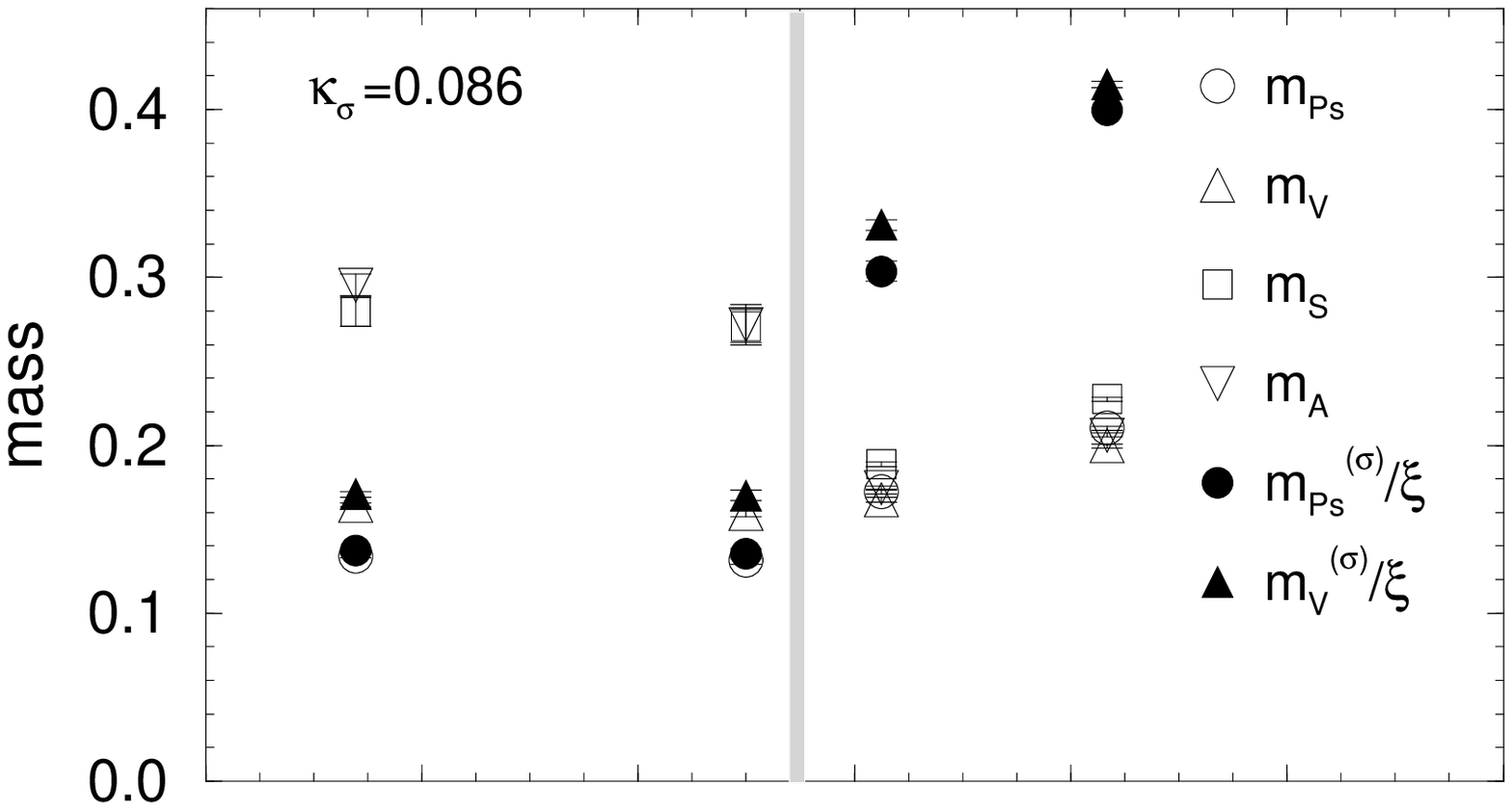,width=\figwidth}
\vspace{-0.9cm} \\
\hspace*{0.05cm}
\leavevmode\psfig{file=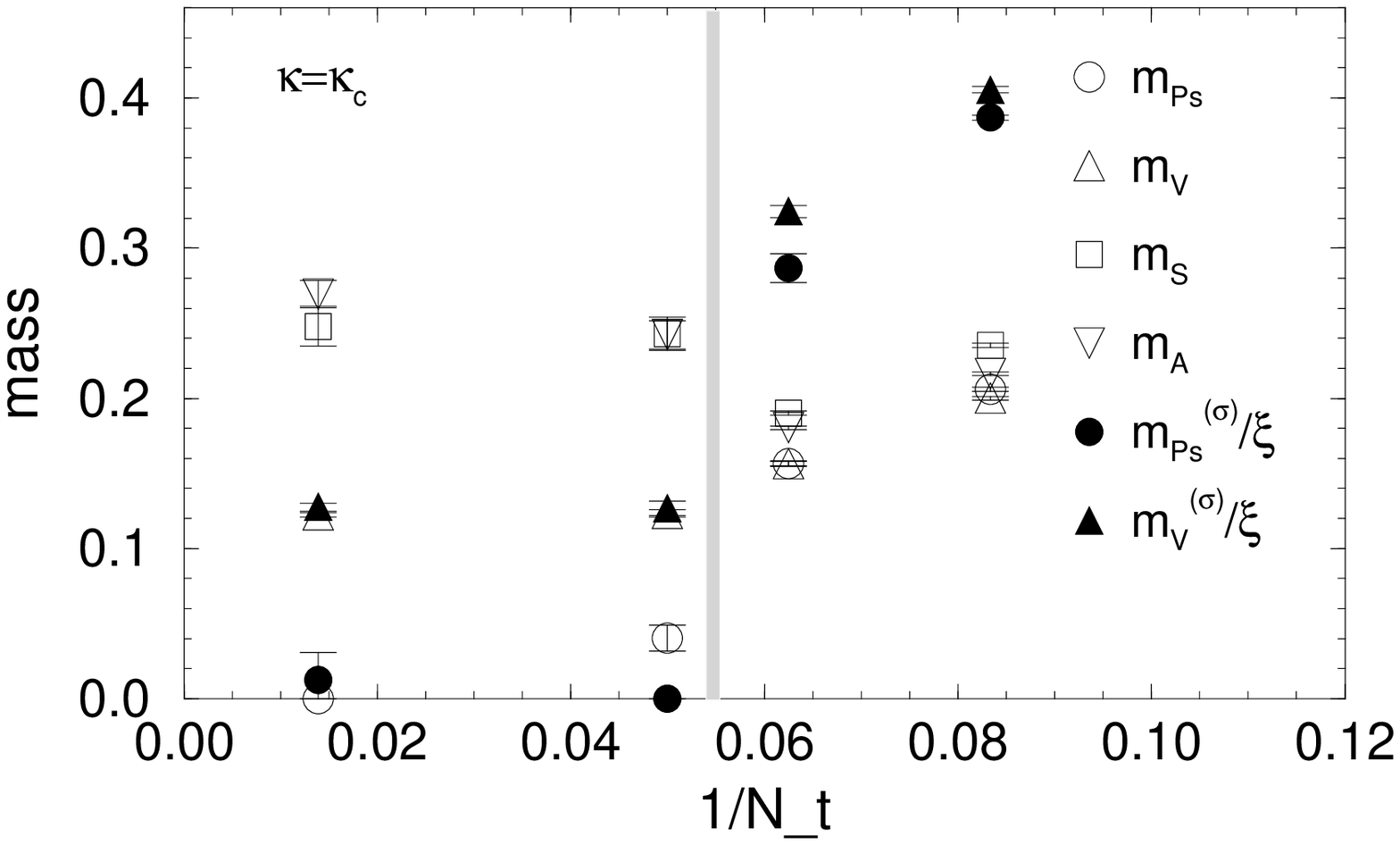,width=1.005\figwidth}
}
\vspace*{0.1cm}
\caption{
Temperature dependence of the mass $m$ (open symbols)
and screening mass $m^{(\sigma)}$ (full symbols)
in units $a_{\tau}^{-1}$, for {\it Set-B},
$\kappa_{\sigma}=0.086$ (upper plot)  and in the chiral limit (lower plot).
The vertical gray lines indicate $T_c$. The data correspond to
$T \simeq 0,~0.93T_c,~1.15T_c$ and $1.5T_c$.  }
\label{fig:t-dep}
%\vspace{-0.5cm}
\end{figure}

\begin{figure}[tb]
\vspace*{-0.6cm}
\center{\hspace*{0.5cm}
\leavevmode\psfig{file=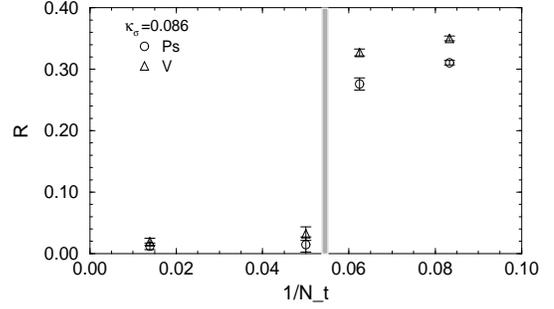,width=\figwidth,height=0.7\figheight}
}
\vspace*{0.5cm}
\caption{
Temperature dependence of $R$, eq. (\ref{e.R}).
The vertical gray line indicates $T_c$. The data correspond to
$T \simeq 0,~0.93T_c,~1.15T_c$ and $1.5T_c$. }
\label{fig:Rt-dep}
\vspace{0.3cm}
\end{figure}

{\it c) Masses and screening masses.}  On the {\it Set-B} data
we fit the $Ps$, $V$, $S$  and $A$
time-correlators to  single hyperbolic functions
at the largest 3 $t$-values \cite{sav}.
{\it Above $T_c$ we assume that
these masses 
characterize the dominant
low energy structure selected by our source}
(e.g.,  the putative, unstable states
described by the wave functions discussed above). 
 We also extract screening masses from spatial correlators
at the largest 3 $x$.
Since the spatial physical distance  is
 large we use {\it point-point} source
for all $T$
(a gauge invariant extended source leads to similar results).
The results for $m$ and $m^{(\sigma)}$ at $\kappa_{\sigma}=0.086$ are shown
 in Fig.~\ref{fig:t-dep} ($m$, and to a smaller extent also $m^{(\sigma)}$
may be  overestimated).
We extrapolate $m$ and $m^{(\sigma)}$
 in $1/\kappa = 2m_0+8$ to the
chiral limit from the 3 quark masses analyzed \cite{extrap}.
Up to $T_c$ screening masses and  masses remain
similar, while above $T_c$
the former become  much larger than the latter,
both at finite quark mass and in the chiral limit 
-- see Fig.~\ref{fig:t-dep} \cite{HK94}.
In the  high $T$ regime of QGP one expects
$m^{(\sigma)} \sim 2\pi/N_{\tau}\ \sim 0.4 (0.5)$ in units of
$a_{\tau}^{-1}$ for $T \simeq 1.15T_c(1.5T_c)$, respectively, 
to be compared with the values in 
Fig.~\ref{fig:t-dep} of $\sim 0.3(0.4)$.
 We introduce:
\vspace{-0.1cm}
\be
R = \frac{ m^{(\sigma)} - m }{ m^{(\sigma)} + m }\
\stb \ 1 - \frac{2m_q }{\pi T}+..,\
m_q  \ll T\ll a_{\tau}^{-1} .
\vspace{-1.2cm}
\label{e.R}
\ee
\no  as a phenomenological parameter to succinctly quantify this
behaviour. Since  at high $T$ the quarks are expected
to exhibit an effective
mass $m_q^{eff} \sim gT/\sqrt{6}$ \cite{fea1}, $R \sim 1 - 0.26g$
and thus can serve as indicator
for how near we are to the hight $T$, perturbative regime. See  Fig.~\ref{fig:Rt-dep}.

{\it In conclusion}, in this
 quenched QCD analysis the changes of the meson properties with temperature
appear to be small below $T_c$, while above $T_c$ they become important
and rapid, but not abrupt. Here we observe apparently
opposing features:
On the one hand, the behaviour of the propagators, in particular
the change in the
ordering of the mass splittings could be accounted for by 
free quark propagation \cite{inver} in the mesonic channels above $T_c$,
which would  also explain
the variation of $m(t)$.
 On the other hand, the behaviour of the wave functions obtained from the 
  4-point correlators
suggests that there can be
 low energy excitations in the mesonic channels   above $T_c$, remnants
of the mesons below $T_c$. They would be characterized by a mass 
giving the location of the corresponding bump in the
spectral function. 
The variation of $m(t)$ with $t$ and with the source
would then indicate a resonance width increasing with $T$,
although it may simply reflect the uncertainty in our treatment of the
low energy states   \cite{sourf}.
Remember that our source is not
chosen arbitrarily but such as to reproduce a ``pure" meson source at $T=0$;
% at high $T$, however, it may not succeed to prevent admixture of
% other contributions, 
% therefore $m$ becomes increasingly ambiguous. 
at high $T$, however, it may allow admixture of
other contributions, 
and $m$ becomes increasingly ambiguous. 
We see chiral symmetry restoration
above $T_c$  both in the masses and in the screening masses,
with the latter
 increasing faster than the former and
remaining below the free gas limit at $T\simeq 1.5 T_c$.
The exact amount of splitting among the channels and the
precise ratio between $m$ and $m^{(\sigma)}$ may, however,
 be affected also by uncertainties in our $\xi$ calibration. Finally, note
 that this is a quenched simulation,
with incomplete dynamics.

A possible physical picture is this: Mesonic excitations subsist above $T_c$
(up to at least $1.5 T_c$) as unstable modes (resonances), in interaction
with unbound quarks and gluons. Our results
are consistent with this, but there may be also other possibilities
(cf \cite{dTar,fea2}, cf \cite{hild} and references therein).
E.g., in  a study of meson propagators including 
 dynamical quarks
but  without  wavefunction information \cite{fea2},
one found masses and (spatial) screening masses $\propto T$
above $T_c$ 
and indication for QGP with ``deconfined, but strongly
interacting quarks and gluons". 
The complex, 
non-perturbative structure of QGP (already indicated by 
 equation of state studies up to far above $T_c$ \cite{eqs}) 
is also confirmed by our analysis
of general mesonic correlators.
From our study however,
the detailed low
energy structure of the mesonic channels appears to present
further interesting, yet unsolved aspects.

Further work is needed to remove
the uncertainties still affecting our analysis. This concerns
particularly the $\xi$ calibration and the
question of the definition of hadron operators at high $T$,
which appear to have been the major deficiencies, besides the smaller lattices,
affecting earlier results \cite{hns1}.
We shall also try to extract
information directly about the
spectral functions \cite{hsp}.
\par\bigskip

\no {\bf Acknowledgments:} We thank
JSPS, DFG and the European Network ``Finite Temperature Phase
Transitions in Particle Physics" for support. H.M. thanks T. Kunihiro 
and I.O.S. thanks F. Karsch for
interesting discussions. We thank F. Karsch for reading the manuscript. 
We are indebted to two anonymous referees for
useful comments.
The calculations have been done on AP1000 at Fujitsu Parallel
Comp. Res. Facilities and Intel Paragon at INSAM, Hiroshima Univ.

\clearpage

\begin{table}[tbh]
%\begin{center}
\begin{tabular}{cccccccc}
set & nr.conf.& $\kappa_{\sigma}$  & $\gamma_F$ & $m_{Ps}$
 & $m_{V}$ & $a$
%(eq. \ref{e.sour})
& $p$
%(eq. \ref{e.sour})
\\
\hline
A& 20 & 0.068 & 5.4  & 0.109(1) & 0.132(2) &
  0.442 & 1.298  \\
\hline
B & 60 & 0.081 & 4.05 & 0.178(1) & 0.196(1) &
  0.379 & 1.289  \\
  & 60 & 0.084 & 3.89 & 0.149(1) & 0.175(1) &
  0.380 & 1.277  \\
  & 60 & 0.086 & 3.78 & 0.134(1) & 0.164(1) &
  0.380 & 1.263  \\
\end{tabular}
%\end{center}
\vspace*{0.2cm}
\caption{Simulation parameters used at every $T$ and 
 meson masses at $T \simeq 0$ (in units
$a_{\tau}^{-1}$). The source
parameters $a,\ p$ eq. (\ref{e.sour}) are extracted from
the $T \simeq 0$ wave function.}
\label{tab:params}
%\vspace{-0.5cm}
\end{table}


\begin{thebibliography}{9}
\bibitem{sav}
For the $S$-meson we only use the connected
correlator.
\bibitem{extrap}
 We assume that the $Ps$ extrapolates in mass squared
for $T < T_c$. Above $T_c$ the $\kappa$ dependence is very weak.
 See \cite{qcdt98}.
\bibitem{inver} We mean $S_0$ with some effective $m_q$ induced by
the interaction.
The larger $m(t)$ in
$Ps$ channel
 as compared with $V$ is a peculiarity of the free Wilson quarks
(\ref{e.fact}). Its
counterpart for staggered fermions is an oscillating behaviour,
 quoted in \cite{fea2} also as indicator of free quark propagation.
\bibitem{sourf} We remark that 
using free quark 
propagators the source dependence is much stronger.
\end{thebibliography}
\end{document}